# A LOCATION-BASED MOVIE RECOMMENDER SYSTEM USING COLLABORATIVE FILTERING

Kasra Madadipouya

Department of Computing and Science, Asia Pacific University of Technology & Innovation

## ABSTRACT

*Available recommender systems mostly provide recommendations based on the users' preferences by utilizing traditional methods such as collaborative filtering which only relies on the similarities between users and items. However, collaborative filtering might lead to provide poor recommendation because it does not rely on other useful available data such as users' locations and hence the accuracy of the recommendations could be very low and inefficient. This could be very obvious in the systems that locations would affect users' preferences highly such as movie recommender systems. In this paper a new location-based movie recommender system based on the collaborative filtering is introduced for enhancing the accuracy and the quality of recommendations. In this approach, users' locations have been utilized and take in consideration in the entire processing of the recommendations and peer selections. The potential of the proposed approach in providing novel and better quality recommendations have been discussed through experiments in real datasets.*

## 1. INTRODUCTION

In traditional recommender systems the recommendation procedure is done with considering about users' past rating history and items features (tagging) and basically no contextual information is taken into account for generating recommendations [1]. However, there are usually various factors influencing users' decision on in reality. Besides, users' demand might change with context as well (e.g. time, location and weather etc) which cannot be accomplished through tradition recommender systems.

As a result just having information about users, items and/or few contextual parameters are not sufficient enough to give accurate real-time emotional-based recommendations, however, more information is necessary in order to provide items to users under certain conditions [2]. For instance, adding the temporal, day/night context, in recommender systems would provide more accurate recommendations which also might reflect user's emotions as well. Additionally, in the personalized content delivery for a recommender system, this fact is crucial to determine whether particular content should be delivered to users or not and if it has to be recommended to users when it should be recommended [2].

With the creation of the new generation recommender system, contextual information has become one of the most valuable knowledge sources to improve recommendations and provide more user specific recommendations under that similar circumstance (i.e., contexts) are related with similar user preferences [1][3] [4].

A better Recommendation System (RS) is the one which delivers recommendations that best match with users' preferences, needs and hopes at the right moment, in the right place and on the right media. This can't be achieved without designing a RS that takes into account all information and parameters that influence user's ratings. This information may concern demographic data [5] preferences about user's domain of interest, quality and delivery requirements as well as the time of the interaction, the location, the media, the cognitive status of the user and his availability. Hence, giving precise recommendations to user is deepened on the relevant gathered information about the user's preferences such as user's location and other parameters.

The rest of the paper is organized as follows; in the next section a review regarding different movie recommender systems are provided. In section three the proposed algorithm is explained in detail which is based on the collaborative filtering and Pearson Correlation Coefficient. Then in the next segment, the results of the experiment is analyzed and presented. Finally, in the last section, conclusions and future enhancement are discussed.

## 2. RELATED WORK

In recommender systems variety of research had been conducted to deliver the most relevant content to user [6][7] [8] in different recommendation scopes for users, such as music recommendation [9][10][11], location-based services [12][13] or even recommendation system for tourism [14][15][16].

In movie recommender system some of the works focus on inputting rich and detailed meta-data (genre and other features) to the system in order to enhance the quality of recommendations. In a movie recommender system by [17], this facility was provided to allow user to even set new keywords for movies which resulted in providing more accurate recommendations based on features which set users. The big advantage of this solution is that it adapts to changes regarding what the users find important, something which can change over time. However, manually categorization of content can be expensive, time-consuming, error-prone and highly subjective. Due to this, many systems aim to automate the procedure.

Another approach of providing high quality recommendation can be achieved by utilizing contextual-information [18] such as time of the day, users' mood, etc.  According to [19] most of the users who are located in the same geographical location have the similar taste to each other. For instance, users who are located in Florida mostly like to watch "Fantasy", "Animation" types of the movie however, the user from Minnesota State are mostly interested to "War", "Drama" type of the movie. Therefore, in recommending items the location of the user is not considered in any steps, as a result it is possible for a user who is looking for a resultant in Chicago get recommendations about some restaurants that are located in other areas such as Seattle [19].
Various algorithms examine by researchers in order to enhance the traditional recommender systems. For instance, in [20], authors presented a novel hybrid approach for movie recommender systems whereby items are recommended based on the collaborative between users as well as analysing contents. In another work by Bogers [21], an algorithm based on Markov random walk proposed which applied on a movie dataset by considering regarding some contextual information. In addition, Rendle et al. [22] applied matrix factorization in a movie dataset with the aim of enhancing the quality and accuracy of the recommendation as well as reducing the complexity of building their model. Simon Funk proposed Regularized singular value decomposition (RSVD) to enhance the accuracy of the recommendation which reduces the dimensionality of the original rating matrix [23].

# 3. THE PROPOSED ALGORITHM

The new method was made to reach to numbers of goals and objectives. Firstly, by contrast of the baseline collaborative filtering that does not take into account users' locations in providing recommendations, in this framework current user location is considered in the system in which user gets different recommendations based on the location. This reflects the fact that if the user changes his/her location the received recommended items will be affected because of changing in the location. Secondly, user preferences that lead him/her to select an item, and correspondingly potentially generates better recommendations based on the user location which increases the locality of the recommended items. Lastly, the main goal is to improve the locality of the recommendations in both peers selection stage and item recommendations with gathering some information about user location and considering about user location in all recommendation processes.

The only recommendation method that is used in the framework contains the following steps as demonstrated in below,

1- Keeping a copy of the user profile in the memory (memory-based approach).
2- Applying similarity index function for the user to find the peers (Pearson Correlation Coefficient).
3- Adding α parameter for those peers who are in the same place with the active user.
4- Applying some filtrations on selected peers to reduce and remove non-related peers.
5- Selecting items for providing recommendations to the active user.
6- Showing the recommended items to the active user.

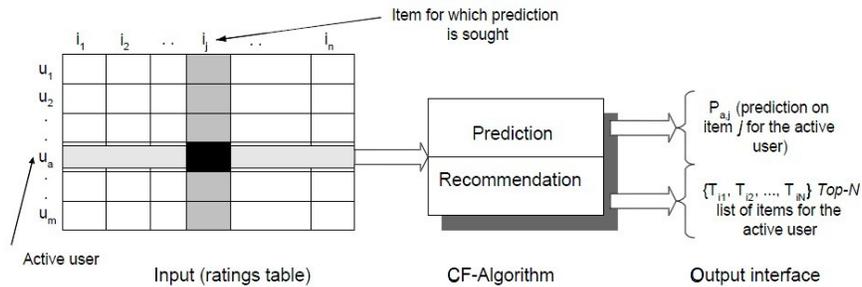

Figure_1 (Overview of collaborative filtering process)

$$sim(a, b) = \frac{\sum_{p \in P}(r_{a,p} - \overline{r_a})(r_{b,p} - \overline{r_b})}{\sqrt{\sum_{p \in P}(r_{a,p} - \overline{r_a})^2}\sqrt{\sum_{p \in P}(r_{b,p} - \overline{r_b})^2}}$$

Figure_2 (Pearson Correlation)

Since, the main algorithm (Figure_2) just relies on one parameter (similarity index) to select peers to the active user and seems all the users with no difference, in this framework we introduced (α) parameters refers to the priority index which directly affects the outcome of the similarity index function that returns a value between (-1 to 1). In prioritize the users those users who are located in the same geographical place with the active user get higher priority with add the constant value (α) to the result of similarity index.

As it depicts in Figure_1 the framework is based on the collaborative filtering in which first the users necessary information about the active user will be stored in different database from the user profile in memory since some temporary analysis and changes might be applied in the user profile which will not be overwritten on the user profile or change it. After copying it, the user profile will pass through from peers selection stage, in that stage, firstly, similarity function is applied for the active user against other users of the system to find which user has the similar taste to the current user. Then peers are selected for the active user. In other word, users will be classified in different groups which each group has the users with the same taste. After that, peers location will be compared with the active user and those who are located in the same geographical location with the active user, their similarity indexes will be increased by adding constant value which is known as α. Then some filtrations will be applied to the peers who are called threshold-based selection, according to which users whose similarity exceeds a certain threshold value are considered as neighbors of the target user [24] and be selected as final peers to give recommendations to the active user based on them. Hence, in this stage all peers who have less than 0.5 similarity index (threshold) will be aborted and will not considered in further steps. After applying filtration to the users in different peers groups, top peers of the active user will be nominated for providing recommendations. Finally, top rated items of the peers who gained the highest similarity indexes will be selected for recommendations. In the final stage, those recommended and selected items are shown to the active users as recommendations.

## 4. EXPERIMENTAL RESULTS

For experimental test, LDOS-CoMoDa [25] dataset which is provided by Laboratorij za digitalno in French is utilized for testing the framework against the baseline collaborative (user-based) filtering approach and therefore, necessary elements from the dataset are extracted for experiment. In the dataset, information about the users is available as part of the user profile and in another section of the dataset list of the rated items, available items and many other parameters are saved. All information in the dataset was gathered by crawling some agencies and was from real user involvement in the system. Hence, there was no direct user involvement in the testing system. However, for implementing the system in the real world all above elements are necessary and should be available for completeness of the system.

Overall the dataset contains 2296 records (rows) which are recorded 120 users' behaviors. The total fields of dataset after modification is six fields, which are userID, itemID, Country (User location), Rating, item description, Related to (Country of item). The rating range is from 1 to 5 which one expresses that the user is highly dislike the item and five represents that the user is highly interested in the item. Number of ratings for each user is various and is not dependent to any parameters such as user location or anything else. The reason for that, it to make the dataset to what happens in reality which the number of ratings items for each user could be different and it is not dependent to any obvious parameter. Therefore, this dataset is obtained by gathering data from different agencies and hence, it is exactly recorded users' behaviors. Table_1 demonstrates all mentioned information about the dataset in summery.

| Number of users | 120 |
|---|---|
| Number of fields | 6 |
| Number of countries | 5 |
| Highest user rate | 5 |
| Lowest user rate | 1 |
| Average rating | 3 |
| Total records of dataset | 2296 |

Table_1 (Dataset information)

The users of the dataset are from five different countries (Denmark, France, Germany, UK, and US). Each country contains different number of the users and this parameter is independent parameter, but in average base on the number of total user each country has 24 users. Table_2 demonstrates, the exact number of each user for each mentioned country.

| Country Name | Number of user |
|---|---|
| Denmark | 41 |
| France | 26 |
| Germany | 6 |
| US | 17 |
| UK | 31 |

Table_2 (User distribution in countries-table)

The experiment has been applied for both baseline Pearson Correlation algorithm and the proposed framework in the form of memory-based user-based filtering. In this experiment, the framework is tested against various value of α (0.1, 0.2, 0.3) separately the results are demonstrated in the following section.

By applying the baseline Pearson Correlation in the dataset, the average gained peers locality was 14.10% which means that only 14.10% of peers are located in the same location with the active users whereas the rest of the peers locations are not the same with the active user. Applying the proposed framework with α = 0.1, gives different result which is demonstrate shows 1.20 per cent (15.33% peers locality rate) improvement in comparison with the baseline approach.

By increasing α to 0.2, the improvement in locality rate reached near to 1.73 per cent (15.33% peers locality rate). Finally for the last experiment α value is set to 0.3 and the framework has been tested against the dataset which shows 1.87 percent (15.97% peers locality rate) enhancement in peers selection stage.

Changing α value on the other hand affects the recommendation results as well in terms of the locations that movie produced. In baseline algorithm only 1.46% of items related to the same location as active user, however, by applying α = 0.1, this amount reached to 1.7% and the number increased to 1.88% and 1.99 by applying α = 0.2 and α = 0.3 respectively.

## 5. CONCLUSION AND FUTURE WORK

In this paper a novel framework to improve locality in user-based collaborative filtering for movie recommender system is proposed and experimental results of the evolution of is presented which gathered results from different tests on dataset analyzed and presented. In addition to that, compare and comparison between the baseline algorithm and the proposed approach is provided in details based on the experimental results.

As already mentioned earlier, the dataset which has been used for this work is part of LDOS-CoMoDa dataset that gathered from various movies agencies. Hence, for future work in order to reach to more accurate result and better analysis, the experiment should be performed on other and bigger datasets which contains hundred thousand of records and have different level of sparseness. In addition, the experimental performed under the lab condition and there was no direct user involvement in experiment the framework and method for getting results. Hence, the proposed method could be implemented in the real environment against the real user experiment.

Moreover, the proposed method in this work applied on one class of filtering method which is Pearson correlation coefficient. Thus, various algorithms from both collaborative and content based approaches could be tested against the dataset and the results compared with the proposed framework in order to provide more accurate analysis.